\documentclass[pra,nofootinbib]{revtex4}

\usepackage{graphicx}
\usepackage{amssymb}
\usepackage{amsmath}
\usepackage{amsfonts}
\usepackage{epstopdf}
\usepackage{epsfig}
\usepackage{wrapfig}

\begin{document}

\title{Periodically driven quantum open systems: Tutorial}
\author{Robert Alicki$^{(1)(3)}$, David Gelbwaser-Klimovsky$^{(2)}$, and Gershon Kurizki$%
^{(2)}$}
\affiliation{$^{(1)}$Institute of Theoretical Physics and Astrophysics, University of
Gda\'nsk, Poland\\
$^{(2)}$Weizmann Institute of Science,  76100 Rehovot, Israel%
\\
$^{(3)}$Weston Visiting Professor at the Weizmann Institute of Science}

\begin{abstract}
\end{abstract}

\maketitle

\section{Introduction}
We present a short derivation and discussion of the master equation for an open quantum system weakly coupled to a heat bath and then its generalization to the case of with periodic external driving based on the Floquet theory. Further, a single heat bath is replaced by several ones. We present also the definition of heat currents which satisfies the second law of thermodynamics and apply the general results to a simple model of periodically modulated qubit. The text is based on \cite{Davies:1974} - \cite{Alicki:2006} but contains also new results.

\section{Thermal generators for constant Hamiltonian}
Consider a system and a reservoir (bath), with a "bare" system Hamiltonians $H^{0}$
and the bath Hamiltonian $H_{R}$, interacting via the Hamiltonian $\lambda H_{int}=\lambda S\otimes R$.
Here, $S$ ($R$) is a Hermitian system (reservoir) operator and $\lambda $ is
the coupling strength ( a generalization to more complicated $H_{int}$ is straightforward). We assume also that
\begin{equation}
[\rho_R , H_R] = 0,\ \mathrm{Tr}(\rho_R\, R) =0 .
\label{ass}
\end{equation}

\par
The reduced, system-only dynamics in the interaction picture is defined as a partial trace
\begin{equation}
\rho (t)=\Lambda (t,0)\rho \equiv \mathrm{Tr}_R \bigl(U_{\lambda}(t,0)\rho\otimes\rho_R U_{\lambda}(t,0)^{\dagger}\bigr)
\label{red_dyn}
\end{equation}
where the unitary propagator in the interaction picture is given by the ordered exponential
\begin{equation}
U_{\lambda}(t,0) = \mathcal{T}\exp\Bigl\{\frac{-i\lambda}{\hbar}\int_0^t S(s)\otimes R(s)\,ds\Bigr\}
\label{prop_int}
\end{equation}
where
\begin{equation}
S(t) = e^{(i/\hbar)Ht} S e^{(i/\hbar)Ht} ,\  R(t)= e^{(i/\hbar)H_R t} R e^{-(i/\hbar)H_R t}.
\label{prop_int1}
\end{equation}
Notice, that $S(t)$ is defined  with respect to the renormalized, \emph{%
physical}, $H$ and not $H^{0}$ which can be expressed as
\begin{equation}
H=H^{0}+\lambda ^{2}H_{1}^{\mathrm{corr}}+\cdots .  
\label{eq:H_S}
\end{equation}
The renormalizing terms containing powers of $\lambda $ are
\emph{ Lamb-shift} corrections due to the
interaction with the bath which cancel afterwards the uncompensated term $H-H^0$ which in principle should be also present in (\ref{prop_int}).  The lowest order
(Born) approximation with respect to the coupling constant $\lambda $ yields 
$H_{1}^{\mathrm{corr}}$, while the higher order terms ($\cdots $) require
going beyond the Born approximation. 
\par
A convenient, albeit not used in the rigorous derivations, tool is  a cumulant expansion for the reduced dynamics 
\begin{equation}
\Lambda (t,0)=\exp \sum_{n=1}^{\infty }[\lambda ^{n}K^{(n)}(t)],
\end{equation}%
One finds that $K^{(1)}=0$ and the Born approximation (weak coupling) consists of terminating
the cumulant expansion at $n=2$, whence we denote $K^{(2)}\equiv K$: 
\begin{equation}
\Lambda (t,0)=\exp [\lambda ^{2}K(t)+O(\lambda ^{3})].
\end{equation}%
One obtains%
\begin{equation}
K(t)\rho =\frac{1}{\hbar^2}\int_{0}^{t}ds\int_{0}^{t}duF(s-u)S(s)\rho S(u)^{\dag }+(%
\mathrm{similar\ terms})  
\label{eq:K(t)}
\end{equation}%
where $F(s)= \mathrm{Tr}(\rho _{R}R(s)R)$.  The \emph{similar
terms} in Eq.~(\ref{eq:K(t)}) are of the form $\rho S(s)S(u)^{\dagger }$ and $S(s)S(u)^{\dagger }\rho $.
\par
The Markov approximation (in the interaction picture) means in all our cases that for long enough time one can use the following approximation
\begin{equation}
K(t)\simeq t\mathcal{L}  \label{eq:L}
\end{equation}
where $\mathcal{L}$ is a Linblad-Gorini-Kossakowski-Sudarshan(LGKS)- generator. To find its form we first decompose
$S(t)$ into its Fourier components
\begin{equation}
S(t)=\sum_{\{\omega\} } e^{i\omega t}S_{\omega },  S_{-\omega }= S_{\omega }^{\dagger}
\label{eq:S}
\end{equation}
where the set $\{\omega\}$ contains \emph{Bohr frequencies} of the Hamiltonian 
\begin{equation}
H= \sum_k \epsilon_k |k\rangle\langle k|,\  \omega = \epsilon_k - \epsilon_l .  
\label{Bohr}
\end{equation}
Then we can rewrite the expression (\ref{eq:K(t)}) as
\begin{equation}
K(t)\rho =\frac{1}{\hbar^2}\sum_{\omega ,\omega ^{\prime }}S_{\omega }\rho S_{\omega
^{\prime }}^{\dag }\int_{0}^{t}e^{i(\omega -\omega ^{\prime
})u}du\int_{-u}^{t-u}F(\tau )e^{i\omega \tau }d\tau +(\mathrm{similar}\text{ 
}\mathrm{terms}).  
\label{eq:K2}
\end{equation}
and use two crucial approximations:
\begin{equation}
\int_{0}^{t}e^{i(\omega -\omega ^{\prime })u}du\approx t\delta _{\omega
\omega ^{\prime }}, \  \int_{-u}^{t-u}F(\tau )e^{i\omega \tau }d\tau \approx {G}(\omega
)=\int_{-\infty }^{\infty }F(\tau )e^{i\omega \tau }d\tau \geq 0 .
\label{eq:rep1}
\end{equation}
This makes sense for $t\gg \max \{1/(\omega -\omega ^{\prime })\}$. Applying these two approximation we obtain $K(t)\rho _{S}=(t/\hbar^2)\sum_{\omega
}S_{\omega }\rho _{S}S_{\omega }^{\dag }{G}(\omega )+(\mathrm{similar}$ $%
\mathrm{terms})$, and hence it follows from Eq.~(\ref{eq:L}) that $\mathcal{L}$ is a special case of the LGKS generator.Returning to the Schroedinger picture one obtaines the following Markovian master equation: 
\begin{eqnarray}
\frac{d\rho }{dt} &=&-\frac{i}{\hbar}[H,\rho ]+\mathcal{L}\rho ,  \notag \\
\mathcal{L}\rho  &\equiv &\frac{\lambda ^{2}}{2\hbar^2}\sum_{\{\omega \}}G(\omega
)([S_{\omega },\rho S_{\omega }^{\dagger }]+[S_{\omega }\rho ,S_{\omega
}^{\dagger }])  
\label{Dav}
\end{eqnarray}
Several remarks are in order:

\noindent (i) The absence of off-diagonal terms in Eq.~(\ref{Dav}), compared
to Eq.~(\ref{eq:K2}), is the crucial property of the Davies generator which can be 
interpreted as a coarse-graining in time of fast oscillating terms. It implies also the commutation of $\mathcal{L}$
with the Hamiltonian part $[H ,\cdot]$.

\noindent (ii) The positivity  $G(\omega )\geq 0$ t follows from Bochner's theorem and is a necessary condition for
the complete positivity of the Markovian master equation.

\noindent (iii) The presented derivation showed implicitly that the notion of
\emph{bath's correlation time}, often used in the literature, is not
well-defined -- Markovian behavior involves a rather complicated cooperation
between system and bath dynamics.  In other words, contrary to what is often done in
phenomenological treatments, \emph{one cannot combine arbitrary }$H$%
\emph{'s with a given LGKS generator}. This is particularly important
in the context of thermodynamics of controlled quantum open system, where it is common to assume Markovian
dynamics and apply arbitrary control Hamiltonians. Erroneous derivations of the quantum master equation can easily lead to violation of the laws of thermodynamics.
\par
If the reservoir is a quantum system at thermal equiliubrium state the additional Kubo-Martin-Schwinger (KMS) condition holds
\begin{equation}
G(-\omega) = \exp\Bigl(-\frac{\hbar\omega}{k_B T}\Bigr) G(\omega),  
\label{KMS1}
\end{equation}
where $T$ is the bath's temperature. As a consequence of (\ref{KMS}) the Gibbs state 
\begin{equation}
\rho_{\beta} = Z^{-1} e^{-\beta H}, \ \beta= \frac{1}{k_B T}  
\label{Gibbs}
\end{equation}
ia a stationary solution of (\ref{Dav}). Under mild conditions (e.g : "the only system operators commuting with $H$ and $S$ are scalars") the Gibbs state is a unique stationary state and any initial state relaxes towards equilibrium ("0-th law of thermodynamics"). A convenient parametrization of the corresponding \emph{thermal generator} reads
\begin{equation} 
\mathcal{L}\rho  = \frac{1}{2}\sum_{\{\omega\geq 0\} }\gamma(\omega)\bigl\{([S_{\omega },\rho S_{\omega }^{\dagger }]+[S_{\omega }\rho ,S_{\omega}^{\dagger }]) + e^{-\hbar\beta\omega}([S_{\omega }^{\dagger},\rho S_{\omega }] +[S_{\omega }^{\dagger }\rho ,S_{\omega}])\bigr\} 
\label{Dav_therm}
\end{equation}
where finally 
\begin{equation}
\gamma(\omega)= \frac{\lambda^2}{\hbar^2} \int_{-\infty}^{+\infty} \mathrm{Tr}\bigl(\rho_R\, e^{iH_R t/\hbar}\,R\, e^{-iH_R t/\hbar}R\bigr)\, dt .  
\label{relaxation}
\end{equation}
\section{Thermal generators for periodic driving}
In order to construct models of quantum heat engines or powered refrigerators we have to extend the presented derivations of Markovian master equation to the case of periodically driven systems. Fortunately, we can essentially repeat the previous derivation with the following amendments:
\par
1) The system (physical, renormalized) Hamiltonian is now periodic
\begin{equation}
\label{Ham_per}
H(t)= H(t + \tau),\ U(t,0) \equiv \mathcal{T}\exp\bigl\{-\frac{i}{\hbar}\int_0^t H(s)\,ds\bigr\},
\end{equation}
and the role of constant Hamiltonian is played by $H$ defined as
\begin{equation}
H= \sum_k \epsilon_k |k\rangle\langle k| ,\ U(\tau,0)= e^{-iHt/\hbar} .
\label{eq:Sq1}
\end{equation}
\par
2) The Fourier decomposition (\ref{eq:S}) is replaced by the following one 
\begin{equation}
U(t,0)^{\dagger}\,S\, U(t,0)=\sum_{q\in \mathbf{Z}}\sum_{\{\omega\}} e^{i(\omega+ q\Omega) t}S_{\omega q},  
\label{eq:Sq}
\end{equation}
where $\Omega = 2\pi/\tau$  and  $\{\omega\}= \{\epsilon_k - \epsilon_l\}$.
The decomposition of above follows from the Floquet theory, however for many models we can obtain it directly using the manifest expressions for the propagator $U(t,0)$.
\par
3) The generator in the interaction picture has form:
\begin{equation} 
\mathcal{L}  = \sum_{q\in \mathbf{Z}}\sum_{\{\omega\}}=\mathcal{L}_{\omega q}
\label{Dav_per}
\end{equation}
where
\begin{equation} 
\mathcal{L}_{\omega q}\rho  = \frac{1}{2}\gamma(\omega+q\Omega)\bigl\{([S_{\omega q},\rho S_{\omega q}^{\dagger }]+[S_{\omega q}\rho ,S_{\omega q}^{\dagger }]) + e^{-\hbar\beta(\omega+q\Omega)}([S_{\omega q}^{\dagger},\rho S_{\omega q}] +[S_{\omega q}^{\dagger }\rho ,S_{\omega q}])\bigr\} .
\label{Dav_per1}
\end{equation}
Returning to the Schroedinger picture we obtain the following master equation:
\begin{equation}
{\frac{d\rho(t) }{dt}}= -\frac{i}{\hbar}[H(t), \rho(t)]+\mathcal{L}(t)\rho(t)
t\geq 0,  \label{ME_per2}
\end{equation}
where 
\begin{equation}
\mathcal{L}(t)= \mathcal{L}(t+\tau ) =\mathcal{U}(t,0)\mathcal{L}\mathcal{U}(t,0)^{\dagger },\ \mathcal{U}(t,0)\cdot
= U(t,0)\cdot U(t,0)^{\dagger}.
\label{gen_per} 
\end{equation}
In particular one can represent the solution of (\ref{ME_per2}) in the form
\begin{equation}
\rho(t) = \mathcal{U}(t,0)e^{\mathcal{L}t}\rho(0) \ ,
t\geq 0.  
\label{sol}
\end{equation}
Any state, satisfying $\mathcal{L}\tilde{\rho}= 0$, defines a periodic steady state (limit cycle)
\begin{equation}
\tilde{\rho}(t) = \mathcal{U}(t,0)\tilde{\rho} = \tilde{\rho}(t+\tau) \ ,
t\geq 0.  
\label{persol}
\end{equation}
Finally one should notice that in the case of multiple couplings and multiple heat baths the generator $\mathcal{L}$ can be always represented as an appropriate sum of the terms like (\ref{Dav_therm}).
\subsection{Heat flows and power for periodically driven open systems}
To derive the laws of thermodynamics in the  case of an arbitrarily fast periodic modulation  we start with the proper definition of heat currents which reproduces the II-law as one expects that the coarse-grained dynamics should satisfy it exactly. The constant (interaction picture) generator can be written as a sum over $M$ independent baths contributions, the harmonics $\{q\Omega\}$ and Bohr quasi-frequencies $\{\omega\}$ 
\begin{equation}
\mathcal{L}=\sum_{j=1}^M \sum_{q\in \mathbf{Z}}\sum_{\{\omega\geq 0\}}\mathcal{L}^j_{q\omega},
\label{decomposition}
\end{equation}
where
\begin{equation}
\mathcal{L}^j_{q\omega}\rho = \sum_{k,l\in I_j}\Bigl({\hat{R}}^{(j)}_{kl}(\omega +q\Omega )\left\{ [S_{l}(q,\omega )\rho
,S_{k}(q,\omega )^{\dagger }]+ h.c.]\right\} + {\hat{R}}^{(j)}_{lk}(-\omega - q\Omega )\left\{ [S_{l}(q,\omega )^{\dagger }\rho,S_{k}(q,\omega )]+ h.c.\right\}\Bigr).
\label{decomposition1}
\end{equation}
Here, the notation ${\hat{R}}^{(j)}_{kl}(x)$ reflects the assumption of $M$-independent heat baths which leads to the decomposition of the matrix ${\hat{R}}_{kl}(x)$ into blocks  $\{{\hat{R}}^{(j)}_{kl}, k,l\in I_j\}$, each corresponding to the $j$-th heat bath and satisfying the Kubo-Martin-Schwinger condition in the form
\begin{equation}
{\hat{R}}^{(j)}_{lk}(-x )= e^{-x/k_B T_j}{\hat{R}}^{(j)}_{kl}(x)\ ,\ k,l\in I_j .
\label{KMS}
\end{equation}
Notice, that a single component $\mathcal{L}^j_{q\omega} $ is also a LGKS generator and possesses a Gibbs-like stationary state (stationarity can be proved combining (\ref{decomposition1}), (\ref{KMS1}) and (\ref{Gibbs}))
\begin{equation}
\tilde{\rho}^j_{q\omega} = Z^{-1} \exp\Bigl\{-\frac{\omega + q\Omega}{\omega}\frac{\bar{H}}{k_B T_j}\Bigr\}, 
\label{gibbs_loc}
\end{equation}
The corresponding time-dependent objects satisfy
\begin{equation}
\mathcal{L}^j_{q\omega}(t) \tilde{\rho}^j_{q\omega}(t) = 0, \ \mathcal{L}^j_{q\omega}(t) =\mathcal{U}(t,0)\mathcal{L}^j_{q\omega}\mathcal{U}(t,0)^{\dagger },\   \tilde{\rho}^j_{q\omega}(t) = \mathcal{U}(t,0)\tilde{\rho}^j_{q\omega} = \tilde{\rho}^j_{q\omega}(t+\tau) \  .
\label{inv.state}
\end{equation}
Using the decomposition (\ref{decomposition}),(\ref{decomposition1}) one can define a \emph{local heat current} which corresponds to the exchange of energy $\omega + q\Omega$ with the $j$-th heat bath for any initial state 
\begin{equation}
{\mathcal{J}^j_{q\omega}}(t) = \frac{\omega + q\Omega}{\omega}\mathrm{Tr}\bigl[( \mathcal{L}^j_{q\omega}(t)\rho(t))\bar{H}(t)\bigr],\ \bar{H}(t)=\mathcal{U}(t,0)\bar{H} ,
\label{curr_loc}
\end{equation}
or in the equivalent form
\begin{equation}
{\mathcal{J}^j_{q\omega}}(t) = -k_B T_j\mathrm{Tr}\bigl[( \mathcal{L}^j_{q\omega}(t)\rho(t))\ln \tilde{\rho}^j_{q\omega}(t)\bigr].
\label{curr_loc1}
\end{equation}
\par
The heat current associated with the $j$-th bath is a sum of the corresponding local ones
\begin{equation}
{\mathcal{J}^j}(t) = -k_B T_j\sum_{q\in \mathbf{Z}}\sum_{\{\omega\geq 0\}}\mathrm{Tr}\bigl[( \mathcal{L}^j_{q\omega}(t)\rho(t))\ln \tilde{\rho}^j_{q\omega}(t))\bigr].
\label{curr_per1}
\end{equation}
Using  Spohn's inequality 
\begin{equation}
\mathrm{Tr}\bigl([\mathcal{L}\rho ][\ln \rho - \ln \tilde{\rho}]\bigr) \leq 0
\label{spohn}
\end{equation}
valid for any LGKS generator $\mathcal{L}$ with a stationary state $\tilde{\rho}$,
one obtains the II-law in the form
\begin{equation}
\frac{d}{dt}S(t) - \sum_{j=1}^M \frac{\mathcal{J}^j(t)}{T_j} \geq 0
\label{IIlaw_periodic}
\end{equation}
where $S(t) = -k_B \mathrm{Tr}\bigl(\rho(t)\ln \rho(t)\bigr)$.
\par
The heat currents in the steady state $\tilde{\rho}(t)$ are time-independent and given by
\begin{equation}
\tilde{\mathcal{J}^j} = -k_B T_j\sum_{q\in \mathbf{Z}}\sum_{\{\omega\geq 0\}}\mathrm{Tr}\bigl[( \mathcal{L}^j_{q\omega}\tilde{\rho})\ln \tilde{\rho}^j_{q\omega})\bigr].
\label{curr_st}
\end{equation}
They satisfy the II-law in the form
\begin{equation}
\sum_{j=1}^M \frac{\tilde{\mathcal{J}}^j}{T_j} \leq 0
\label{IIlaw_st}
\end{equation}
while, according to the I-law
\begin{equation}
-\sum_{j=1}^M \tilde{\mathcal{J}^j} = - \tilde{\mathcal{J}} = \bar{\mathcal{P}}
\label{power_st}
\end{equation}
is the averaged power (negative when the system acts as a heat engine). Notice, that in the case of a single heat bath
the heat current is always strictly positive except the case of no-driving when it is equal to zero.
\par
\section{The example. Qubit with the diagonal modulation}

\par
We apply the results of the previous Section to particularly simple model of a qubit with diagonal time-dependent Hamiltonian  
\begin{equation}
H(t) = \frac{1}{2}\omega(t)\sigma^3 \ ,\ \omega(t+\tau)= \omega(t) .
\label{TLS1}
\end{equation}
A qubit is coupled to two independent heat bath (hot and cold) at the temperatures $T^h , T^c$, respectively , by the interaction Hamiltonian
\begin{equation}
H_{int} = \sigma^1\otimes (B^h + B^c) .
\label{TLS2}
\end{equation}
Because of the diagonal form of $H(t)$ the Floquet formalism of above drastically simplifies due to
\begin{equation}
F(t) = F(0) = \exp\bigl\{-i\omega_0\tau \sigma^3/2\bigr\}\ ,\  \omega_0 = \frac{1}{\tau}\int_0^{\tau} \omega(t)dt.
\label{Floquet_TLS}
\end{equation}
The master equation for the density matrix of TLS has form (notice time-independence of the dissipative part)
\begin{equation}
{\frac{d\rho(t) }{dt}}=-i\frac{1}{2}\omega(t)[\sigma^3 , \rho(t)]+\mathcal{L}^c\rho(t) +\mathcal{L}^h\rho(t).
\label{ME_qubit}
\end{equation}
Here, putting $a = c,$ or $a = h$ we have
\begin{equation}
\mathcal{L}^a\rho =\sum_{q\in \mathbf{Z}}\mathcal{L}^a_q\rho \
\label{gen_qubit1}
\end{equation}
where
\begin{equation}
\mathcal{L}^a_q\rho = \frac{P(q)}{2}\Bigl(G^a(\omega_0+ q\Omega)\bigl([\sigma^- \rho, \sigma^+] +[\sigma^-, \rho \sigma^+]\bigr) +
G^a(-\omega_0- q\Omega)\bigl([\sigma^+ \rho, \sigma^-] +[\sigma^+, \rho \sigma^-]\bigr)\Bigr)
\label{gen_qubit2}
\end{equation}
Here
\begin{equation}
G^a(\omega)= \int_{-\infty}^{+\infty} e^{i\omega t}\langle B^a(t)B^a\rangle dt = e^{\omega/k_B T_a} G^a(-\omega) .
\label{spectral_qubit}
\end{equation}
and the probability distribution $P(q)$ is given by
\begin{equation}
P(q) = |\xi(q)|^2\  , \ \xi(q) = \frac{1}{\tau}\int_0^{\tau} e^{i\int_0^t(\omega(s)-\omega_0)ds }e^{iq\Omega t} dt .
\label{probability}
\end{equation}

Notice that the dynamics (\ref{ME_qubit}) transforms the diagonal elements into diagonal ones and the stationary states are also diagonal.
The decomposition of the generators (\ref{gen_qubit2}) into harmonic components leads to the picture of the periodically driven qubit in a steady state as equivalent (in a certain sense) to a qubit with a time-independent Hamiltonian $H_0 =\omega_0\sigma^3/2$
coupled to two sequences of "effective heat-baths" with the "temperatures"
\begin{equation}
T_a(q) =\frac{\omega_0}{\omega_0 - q\Omega} T_a  \ ,\  a=c,h  .
\label{effective_T}
\end{equation}
Notice that the effective temperatures can be positive or negative. However, the equivalence is restricted to the form
of the stationary states $\tilde{\rho}^a_q$ and $\tilde{\rho}$
\begin{equation}
\tilde{\rho}^a_q = Z^{-1}(a,q) \exp\Bigl(-\frac{H_0}{k_B T_a(q)}\Bigr)  \ ,\  \tilde{\rho}= Z^{-1} \exp\Bigl(-\frac{H_0}{k_B T_{\mathrm{eff}}}\Bigr)  .
\label{stat_TLS}
\end{equation}
where 
\begin{equation}
T_{\mathrm{eff}} =\frac{\omega_0}{k_B \ln {\frac{\bar{R}^c_{e}+\bar{R}^h_{e}}{\bar{R}^c_{g}+\bar{R}^h_{g}}}} .
\label{stat_TLS1}
\end{equation}
and the averaged decay rate from the excited state $\bar{R}^a_{e}$ and the excitation rate from the ground state $\bar{R}^a_{g}$ read
\begin{equation}
\bar{R}^a_{e}= \sum_{q\in \mathbf{Z}} P(q) G^a(\omega_0+ q\Omega)\ ,\ \bar{R}^a_{g}= \sum_{q\in \mathbf{Z}} P(q) G^a(-\omega_0- q\Omega).
\label{decay_qubit}
\end{equation}
On the other hand the "single-harmonic current" has form
\begin{equation}
\tilde{\mathcal{J}^a_q} = \frac{T_a}{T_a(q)}\mathrm{Tr}\bigl[( \mathcal{L}^j_q\tilde{\rho})H_0)\bigr]=
\Bigl(1 - \frac{q\Omega}{\omega_0}\Bigr)\mathrm{Tr}\bigl[( \mathcal{L}^j_q\tilde{\rho})H_0)\bigr]
\label{curr_TSL}
\end{equation}
which differs by the factor $\frac{T_a}{T_a(q)}$ from the heat current related to a heat bath coupled to a system with the Hamiltonian $H_0$. Due to this modification the total heat current equal to minus power is generally different from zero in contrast to the case of absent modulation.
\par
For the special case of the modulation in a form of two alternating pulses per period
\begin{equation}
\omega(t) = \omega_0 + \pi \sum_{q\in \mathbf{Z}}\bigl( \delta (t - (q +1/4)) - \delta (t - (q +3/4))\bigr) .
\label{modulation_I}
\end{equation}
we obtain 
\begin{equation}
P_{2n} = 0 \ , P_(2n+1) = \frac{4}{\pi^2} \frac{1}{(2n +1)^2} ,\ n = 0 ,1, 2,... .
\label{prob_I}
\end{equation}
%

%\bibliographystyle{hunsrt}
%\bibliography{bib}

\end{document}